\documentclass[aps,prl,twocolumn,showpacs,groupedaddress,superscriptaddress]{revtex4-1}
\pdfoutput=1
\usepackage{graphicx}
\usepackage{color}

\begin{document}

\title{Graphene supports the propagation of subwavelength optical solitons}

\author{M.~L.~Nesterov}
\author{J. Bravo-Abad}
\affiliation{Departamento de F\'{i}sica Te\'{o}rica de la Materia Condensada, Universidad Aut\'{o}noma de Madrid, E-28049 Madrid, Spain}
\author{A. Yu. Nikitin}
\affiliation{Instituto de Ciencia de Materiales de Arag\'on and Departamento de F\'{i}sica de la Materia Condensada, CSIC-Universidad de Zaragoza, E-50009 Zaragoza, Spain}
\author{F.~J.~Garcia-Vidal}
\email{fj.garcia@uam.es}
\affiliation{Departamento de F\'{i}sica Te\'{o}rica de la Materia Condensada, Universidad Aut\'{o}noma de Madrid, E-28049 Madrid, Spain}
\author{L.~Martin-Moreno}
\affiliation{Instituto de Ciencia de Materiales de Arag\'on and Departamento de F\'{i}sica de la Materia Condensada, CSIC-Universidad de Zaragoza, E-50009 Zaragoza, Spain}%

\date{\today}

\begin{abstract}
We study theoretically nonlinear propagation of light in a graphene monolayer. We show that  the large intrinsic nonlinearity of graphene at optical frequencies enables the formation of quasi one-dimensional self-guided beams (spatial solitons) featuring subwavelength widths at moderate electric-field peak intensities. We also demonstrate a novel class of nonlinear self-confined modes  resulting from the hybridization of surface plasmon polaritons with graphene optical solitons.
\end{abstract}


\maketitle

The experimental discovery and isolation of graphene monolayers from bulk graphite \cite{Novoselov2004} has attracted great interest during the last years.
The study of graphene properties has become a hot topic of research within the physics and nanoscience communities \cite{Geim2009Rew} as it promises, among others, a variety of optical and opto-electronical applications \cite{Ferrari2010Rew,Liu}. Very large values of the nonlinear optical susceptibilities corresponding to multiple harmonic generation were theoretically predicted \cite{Mikhailov2007,Ishikawa2010} and have been experimentally verified very recently in the case of third-order nonlinear effects \cite{Hendry2010}. Still, it is an open question whether this high nonlinear coefficient, which occurs in a two-dimensional (2D) system, could induce strong nonlinear effects in electromagnetic (EM) modes that extend on the three spatial dimensions.

One of the nonlinear effects with greater potential for controlling light propagation at the micro- and nano-scales is the formation of temporal and spatial EM solitons \cite{Boyd1992,Agrawal2003,Kivshar2003,Torner,Zhang2007}. In this Letter we demonstrate that 2D graphene monolayers support spatial non-diffracted beams (i.e., solitons) of subwavelength width in the optical regime. We illustrate this capability by analyzing two arrangements leading to solitons with different polarizations: a graphene monolayer embedded into a conventional dielectric waveguide and a graphene sheet placed on top of a metal-dielectric structure. We analyze in detail the formation of spatial solitons and the relation between soliton width and input power, showing that the subwavelength scale can be reached by using feasible values for the beam peak intensity. We also develop a quasi-analytical model that is able to capture the basic ingredients of the numerical results.

\begin{figure}[htbp]
\includegraphics[width=\columnwidth]{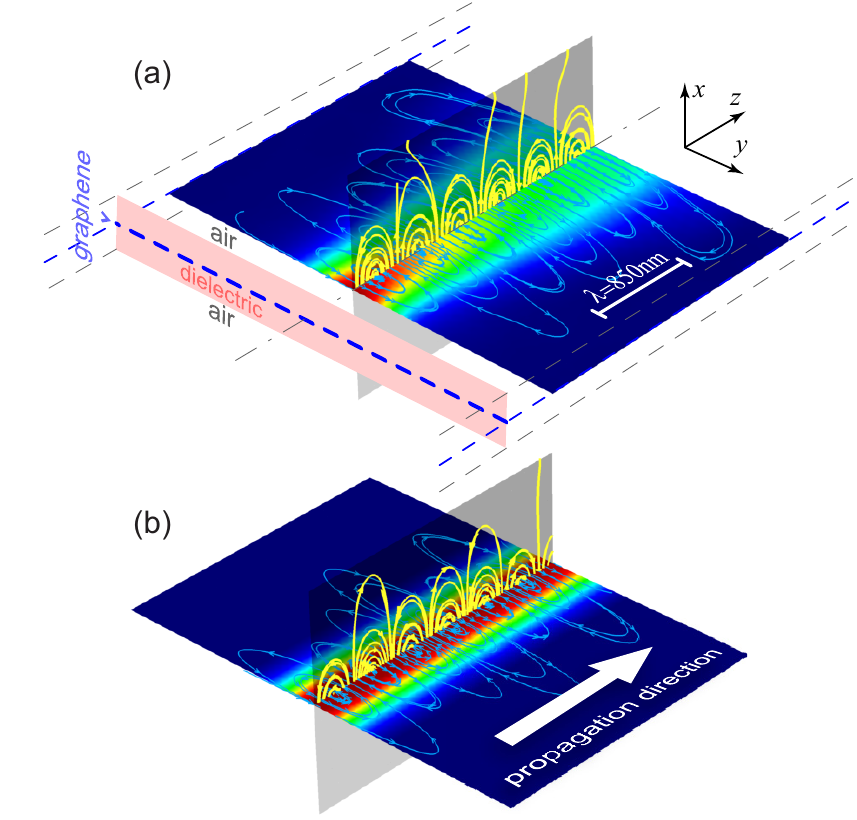}
\caption{\label{fig:Geom} (color online) Geometry and TE soliton formation. An optical beam propagates inside the waveguide with a graphene monolayer located in the center in the low power (a) and high power (b) regimes. Panels (a) and (b) show slices of the beam intensity evaluated at the graphene layer, the yellow lines represent magnetic vector field whereas white lines depict the electric vector field.}
\end{figure}

The first structure in our analysis consists of a single graphene monolayer placed inside a planar linear dielectric waveguide, see Fig.~\ref{fig:Geom}~(a). This dielectric waveguide provides vertical
confinement in the $x$-direction for the propagating EM mode. Graphene must be \textit{physically} considered as a 2D material with nonlinear conductivity. But \textit{mathematically} we can approximate graphene by a very thin layer of a finite thickness introducing an \textit{effective dielectric constant}. Then we can treat graphene using the Maxwell equations for bulk media. We have checked that both approaches give virtually the same numerical results. Since 2D and 3D treatments are equivalent, we take directly the nonlinear susceptibility from the experiment, and approximate graphene by a thin $d_{gr} =0.3$~nm-thick layer. The nonlinear polarization density in graphene is $\mathbf{P}_{\mathrm{NL}}(\mathbf{r}, t) = \epsilon_0\epsilon_{\mathrm{NL}}(\mathbf{r},t)\mathbf{E}_{\mathrm{NL}}(\mathbf{r}, t)$, where $\mathbf{E}_{\mathrm{NL}}(\mathbf{r}, t)$ is the electric field and
$\epsilon_{\mathrm{NL}}(\mathbf{r}, t) =\chi_{gr}^{(3)}[\mathbf{E}_{\mathrm{NL}}(\mathbf{r}, t)]^2$ is the nonlinear contribution to the equivalent permittivity of the graphene layer. The parameter $\epsilon_0$ is the vacuum permittivity. From the polarization density the nonlinear current is obtained as $\mathbf{j}_{\mathrm{NL}}(\mathbf{r}, t) = \partial\mathbf{P}_{\mathrm{NL}}(\mathbf{r}, t)/\partial t$. Throughout this paper, we consider the operating wavelength, $\lambda_0=850$~nm, for which a Kerr-type third-order effective nonlinear susceptibility~$\chi_{gr}^{(3)}\simeq1.5\times10^{-7}$~esu, ($\chi_{gr}^{(3)}\simeq2.095\times10^{-15}$~m$^2/$V$^2$ in SI units) has been measured~\cite{Hendry2010}. The nonlinear eigenmode problem is formulated in terms of the 3D vector Maxwell equations and solved self-consistently in the continuous wave regime using the finite element method \cite{comsol}.

For the case depicted in Fig. 1(a), the initial solution for the iterative method has a form of a TE-polarized beam propagating in the $z$-direction with a gaussian shape along the $y$-direction and a waveguide profile in the $x$-direction. On each step of the iterative process, the EM fields are calculated by solving the propagation problem with the eigenmode solution introduced as a source. In these calculations, we neglect third-order nonlinear effects in the high-index dielectric material surrounding the graphene layer. This assumption is justified by the fact that the magnitude of the third-order nonlinear optical susceptibility in conventional high-index dielectric media is several orders of magnitude smaller than the one characterizing graphene.

Figure 1 illustrates the formation of a non-diffracted beam at an operating wavelength $\lambda_0=850$ nm and for a dielectric waveguide of thickness $300$ nm, characterized by a linear dielectric permittivity, $\epsilon_d=2.25$. When the beam intensity (defined as the modulus of the Poynting vector) at maximum is low ($I<10^{13}$ W/m$^2$), the system operates in the linear regime and the beam diffracts while traveling in this structure, see Fig. 1(a). However, our numerical calculations show that, for high enough intensity ($I>10^{13}$ W/m$^2$), the nonlinearity of graphene can compensate diffraction leading to the formation of EM solitons. This is illustrated in Fig. 1(b), computed for $I=10^{14}$~W/m$^2$, showing a non-diffracted beam with a lateral size of the order of $\lambda_0$.  The soliton field is laterally confined due to the self-induced change of the effective refractive index, similarly to what happens to a beam traveling within a bulk nonlinear waveguide ~\cite{Kivshar2003}. In contrast to a conventional nonlinear waveguide, in which the nonlinear index change occurs in the whole volume, here in our system the 3D beam is laterally self-guided thanks to the nonlinearity that is only present in the 2D graphene sheet.  We stress that these spatial solitons, sustained by a single graphene sheet, are very different to those supported by a metamaterial composed of graphene-dielectric superlattices in the terahertz regime, which propagate perpendicularly to the graphene layers \cite{Biancalana2011}. We also emphasize that the class of bright self-guided solitonic modes observed in Fig.~1(b) could not be supported by a thin metal film. In general, for the frequency range considered in this work, metal films of nanometric thickness feature complex values of the nonlinear third-order susceptibility, $\chi^{(3)}$, such that $\textrm{Re}\chi^{(3)} < 0$ and $\textrm{Im}\chi^{(3)}$ is positive and large \cite{Xenogiannopoulou07}, i.e., they display self-defocusing nonlinearities with large nonlinear absorption losses.

\begin{figure}[tbp]
\includegraphics[width=8.5cm]{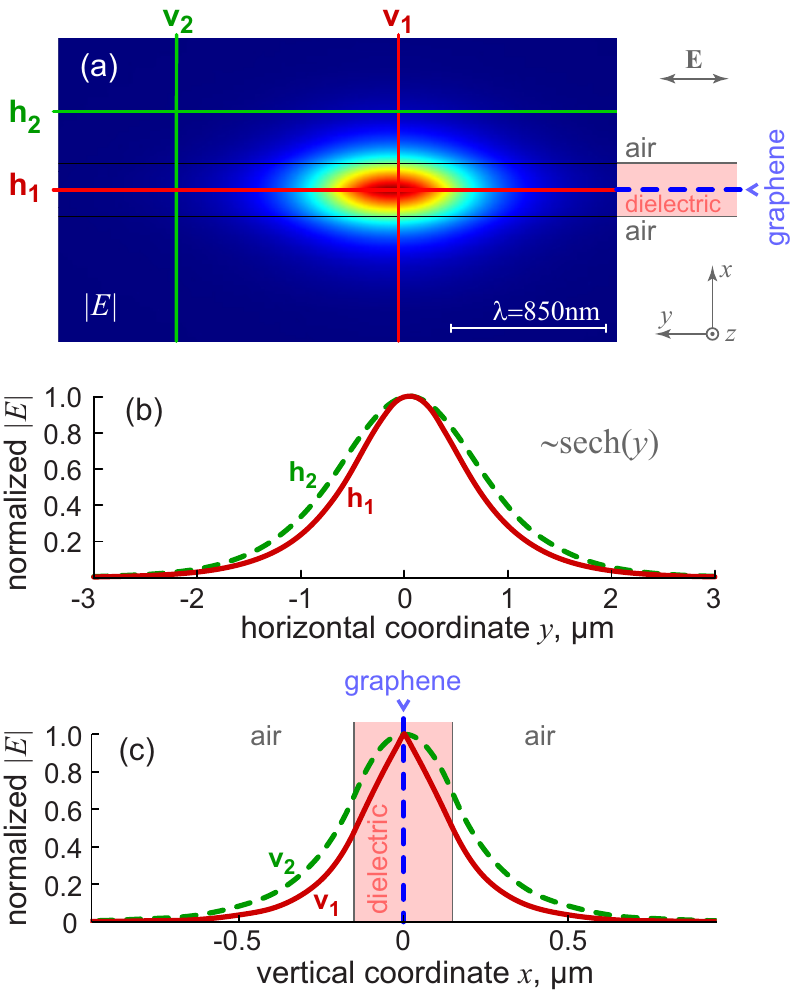}
\caption{\label{fig:eigenmode} (color online) Soliton profile analysis. Panel (a) shows the transversal {\it E}-field distribution, $|E|$, of the soliton mode for the case in which the peak intensity is $I=1.8 \times 10^{14}$W/m$^2$. The horizontal cross-sections $h_1$ and $h_2$ of the normalized {\it E}-field (panel b) display a conventional soliton profile proportional to $\mathrm{sech}(y)$. For the vertical cross-sections (panel c),  the {\it E}-field at $v_2$ has the standard profile of a linear waveguide mode. For the cross-section evaluated at $v_1$ the shape of the beam corresponds to that of a nonlinear system. The filled area on the panel (c) shows the location of the dielectric waveguide.}
\end{figure}

Optical solitons in graphene should be observable with current samples and moderate beam intensities. Although the considered peak intensity in Fig. 1(b) is much higher than the reported damage threshold of graphene for continuous wave excitation, $I_{\mathrm{CWth}}\sim10^{10}$~W/m$^2$ \cite{Smet2009}, it is still well below the damage threshold of graphene for $200$ fs pulses, $I_{\mathrm{Pth}}\sim10^{16}$~W/m$^2$ \cite{Roberts2011}. Our continuous wave description of the soliton propagation under pulsed excitation is fully justified as the optical cycle associated with $\lambda_0=850$ nm is two orders of magnitude shorter than a $200$ fs  pulsed beam.

Additional insight on the physical process can be gained from the study of the beam profile. Figure 2(a) renders the normalized transversal electric field profile $|{\bf E}(x,y)|$ of the calculated 3D eigenmode for $I=1.8\times10^{14}$ W/m$^2$ (high power regime). The field cross-section along the $y$-direction [see Fig. 2(b)] can be accurately fitted by the function ${sech}(y/w(x))$, where $w(x)$ is a measure of the lateral beam size, which slightly depends on $x$. This functional form for graphene EM solitons will be discussed later on. On the other hand, the confinement of the $E$-field along the $x$-direction is governed by the total internal reflection at the boundaries of the dielectric waveguide. The normalized $E$-field cross-section along the $x$-direction changes as the $y$ coordinate is varied (see Fig.~\ref{fig:eigenmode}(c)). This change is more significant than in planar nonlinear waveguides and represents a distinct manifestation of the unusually large nonlinear optical current supported by the 2D graphene sheet that, in the present case, substantially exceeds the linear one.

We turn now to analyze the dependence of soliton width (characterized by the full width at half maximum (FWHM), $a$, of the soliton {\it E}-field) on the external intensity illuminating the system. In order to do this, we have computed the nonlinear eigenmodes for several peak $E$-field amplitudes in the graphene layer. The results, in terms of the corresponding intensity distributions (which in each case have been normalized to the maximum beam intensity), are summarized in the inset of Fig.~3. As the maximum value of the $E$-field in the graphene layer ($|E|_{max}$) is increased from $0.8\times 10^8$ V/m (bottom panel) to $4.2\times 10^8$ V/m (top panel), $a$ decreases from $a$=2 $\mu$m (more than 2 times the wavelength of the external illumination) to $a$=0.253  $\mu$m  (well inside the subwavelength regime). The results displayed in Fig. 3 represent a novel instance, in a strict 2D system, on how the balance between nonlinearity and diffraction can yield self-guided propagating beams with subwavelength lateral confinement. In this context, it is important to point out that when graphene losses are incorporated into the calculations (these losses stem from the linear part of the graphene conductivity), the propagation length $L$ of the soliton, defined as $L=1/2\mathrm{Im}(\beta_{NL})$, $\beta_{NL}$ being the complex propagation constant of the nonlinear mode, is barely dependent on $a$. In fact, we have found numerically that $L$ varies between $15$ and $20$ $\mu$m for all the soliton widths considered in this work. This independence of the propagation length on the field confinement is very different to what is observed in other subwavelength-confined EM modes as, for example, surface plasmon polaritons.

\begin{figure}[htbp]
\includegraphics[width=8.5cm]{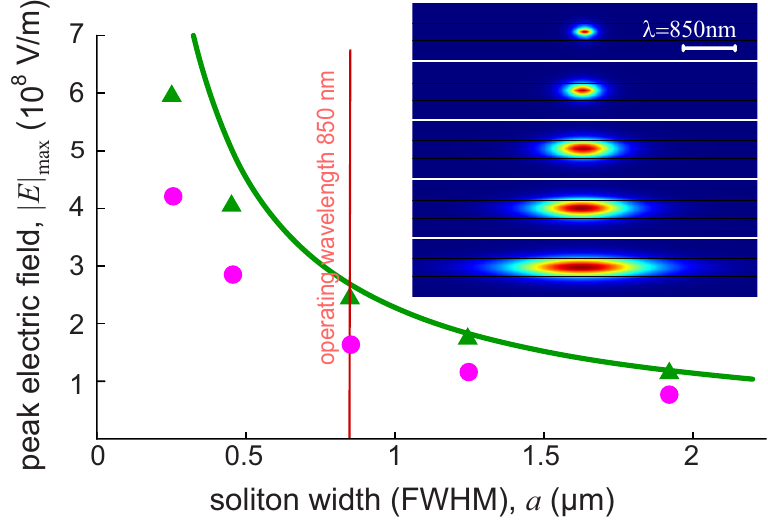}
\caption{\label{fig:power} (color online) Dependence of the soliton width with the input power. The inset shows the normalized intensity distributions for decreasing values of the input power (from top to bottom), resulting in spatial solitons of increased width. The operating wavelength in all cases is $\lambda_0=850$~nm. Main panel presents the peak electric field dependence with the soliton width. Circular dots represent the values obtained with the full nonlinear calculation whereas the solid line renders the results from the quasi-analytical treatment (see main text). When the profile $\hat{{\bf A}}(x)$ and the propagation constant $\beta$ corresponding to the numerical calculation is included (triangular dots), the agreement between analytics and numerics is improved.}
\end{figure}

To account for the physical origin of the above described dependence of the soliton width on the peak electric field amplitude, we have adapted to this problem the theoretical approaches used to describe soliton formation in conventional 3D nonlinear optical materials \cite{Boyd1992, Agrawal2003,Kivshar2003}. For this quasi-analytical treatment, we employ the 3D modeling of the graphene layer, which, as mentioned before, gives virtually the same results as a description based on a strictly 2D conductivity. Within this approach the propagation of light  {\emph {inside}} the graphene layer is formulated in terms of the non-homogeneous vector Helmholtz's equation,
\begin{equation}
c^2 \epsilon_0 \left[ \left(\frac{n_s}{c}\right)^2 \frac{\partial^2}{\partial t^2}-\nabla^2 \right]\: {\bf A}({\bf r},t)={\bf j}_{NL}({\bf r},t)
\end{equation}
where ${\bf A}({\bf r},t)$ is the magnetic potential vector (i.e., \mbox{${\bf A}({\bf r},t)=-\partial {\bf E}({\bf r},t) / \partial  t$}, choosing the gauge ${\bf \nabla} \cdot {\bf A}=0$) and $n_s$ is the linear refractive index of graphene.
To solve Eq. (1), we start by assuming that its solutions are of the form
\begin{equation}
{\bf A}({\bf r},t)=\frac{1}{2}[\hat{{\bf A}}(x) \: F(z,y) \: {\textrm {exp} }[i(\beta z-\omega t)]+c.c.]
\end{equation}
where $\hat{{\bf A}}(x)$ is, in principle, an arbitrary function that governs the confinement of the EM field along the $x$-direction (see definition of axes in Fig.~1).  As deduced from Eq.~(2), $\hat{{\bf A}}(x)$ also defines the polarization of the considered modal profile. The EM field profile in the graphene plane is controlled by the complex function $F(z,y)$, whereas the corresponding propagation constant along the $z$-direction is given by $\beta$.

Now, we insert Eq.~(2) into Eq.~(1), we apply the slowly varying amplitude approximation, and we project the left and right-hand-side of the resulting equation over $[\hat{{\bf A}}(x)]^{*T}$ (where $[\:]^{*T}$ stands for the transpose conjugate). Then, we define the auxiliary function \mbox{$f(z,y) \equiv F(z,y)\exp(-i \phi z)$}  (where $\phi \equiv (k_s^2-\beta^2+I_2/I_1)/2$, $k_s=n_s^2 \omega^2/c^2$, $I_1 \equiv \int_{-\infty}^{\infty} dx |\hat{{\bf A}}(x)|^2$ and $I_2 \equiv \int_{-\infty}^{\infty} dx [\hat{{\bf A}}(x)]^{*T} \partial^2 \hat{{\bf A}}(x)/\partial x^2$). Using these definitions, after some algebra, one finds that Eq.~(1) can be rewritten in terms of the function $f(z,y)$ as
\begin{equation}\label{n2}
    2i\beta\frac{\partial f(z,y)}{\partial z}+\frac{\partial^2 f(z,y)}{\partial y^2}+g|f(z,y)|^2 f(z,y)=0
\end{equation}
where $g \equiv \frac{3}{4}\omega^4\chi^{(3)}_{gr} I_3/I_1c^2$, with $I_3 \equiv \int_{-d_{gr}/2}^{+d_{gr}/2} dx |\hat{{\bf A}} (x)|^4$. The crucial point to realize is that Eq.~(3) corresponds to the standard form of the \emph{nonlinear Schrodinger equation}, whose solutions have  a canonical first-order soliton form \cite{Boyd1992, Kivshar2003},
\begin{equation}\label{n3}
 f(y,z)=\frac{1}{w}\sqrt{\frac{2}{g}}\, \mathrm{sech}(y/w)\exp(iz/2\beta w^2)
\end{equation}
 where $w$ is the conventional definition of the soliton width, which in terms of the soliton FWHM is given by $w \approx a/2.64$. Physically, Eq.~(3) and its corresponding solution given in Eq.~(4) can be interpreted as those governing the propagation of light in a special class of index-guided waveguide in which the refractive index contrast between the core and the cladding is induced by the intensity of the propagating beam itself. Importantly, Eq.~(4) confirms the existence of soliton solutions in graphene, as observed in the numerical experiments reported in Figs.~(1)--(3).  Notice that the strength of the effective nonlinearity is characterized by the parameter $g$, which is proportional to both  $\chi^{(3)}_{gr}$ (related to the intrinsic nonlinearity of graphene in a free-standing configuration) and $I_3/I_1$, which provides a measure of the fraction of EM energy that flows inside the graphene sheet.

Inspired by the theoretical approaches used traditionally in nonlinear optics \cite{Boyd1992, Agrawal2003,Kivshar2003}, we assume that both the vector function $\hat{{\bf A}}(x)$ and the propagation constant $\beta$ of the modal profile correspond to those obtained numerically for the linear counterpart of the structure sketched in Fig.~1(a). The results computed within this approximation are displayed in Fig.~3 (see solid line), showing a qualitative agreement between the analytical results and the full numerical calculations. We emphasize that no fitting parameters are used in this comparison. The discrepancy between analytics and full numerics becomes larger as the value of $a$ decreases. This fact can be ascribed to the difference between the profile $\hat{{\bf A}}(x)$ obtained for the linear case and that computed numerically for the full nonlinear problem, which increases as $a$ decreases. This point is confirmed by the additional results displayed in Fig.~3 (triangular points), which show how the agreement between analytics and numerics improves when we introduce in Eq.~(2) both the self-consistent profile, $\hat{{\bf A}}(x)$ (obtained at $y=0$), and the propagation constant $\beta$ corresponding to our nonlinear simulations. The remaining difference can be traced back to the separability in the $x$ and $(y,z)$ coordinates implied by Eq.(2) that cannot fully account for the complexity of the graphene EM solitons.

\begin{figure}
\includegraphics[width=8.5cm]{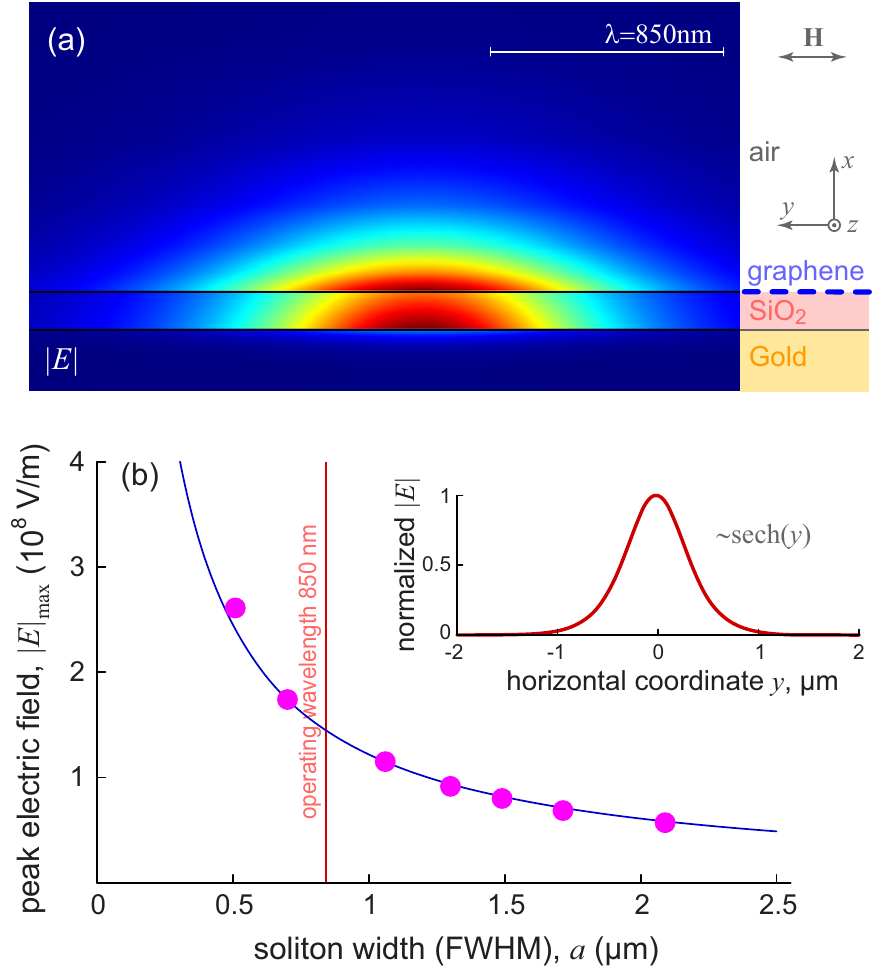}
\caption{\label{fig:SPPsoliton} (color online) TM soliton formation. The geometry consists of the graphene monolayer and gold half-space separated by a silicon dioxide layer of thickness of $100$~nm. The operating wavelength is $\lambda_0=850$ nm. The density plot in panel (a) is the transversal $E$-field distribution, $|E|$, associated with the excitation of a TM SPP-soliton. Panel (b) shows the dependence of the soliton FWHM, $a$, with the input power, measured as the peak {\it E}-field amplitude evaluated at the graphene monolayer. The dots represent the numerical results whereas the solid line is a fitting to a $1/a$ function. Inset  of panel (b) renders an horizontal cross-section of the {\it E}-field amplitude along the $y$-direction at the graphene layer.}
\end{figure}

Finally, we show that TM-polarized optical solitons can also propagate along a graphene monolayer. A graphene structure that is able to support these TM optical solitons is rendered on Fig.~\ref{fig:SPPsoliton}. Here the vertical confinement is provided by a surface plasmon polariton (SPP) mode that is propagating on the interface between gold and a dielectric film. The graphene monolayer, which is characterized by a large nonlinear third-order susceptibility, must be separated from the metal surface by a dielectric spacer. We have chosen a $100$ nm silicon dioxide layer, just for proof-of-principles purposes. Our calculations show that this system supports the propagation of a very peculiar class of TM soliton, which results from the {\emph {hybridization}} between the SPP supported by the metal-dielectric interface and the soliton propagating in the graphene sheet. The computed transversal $E$-field distribution, $|{\bf E}(x,y)|$, of this hybrid SPP-soliton solution is plotted in Fig.~\ref{fig:SPPsoliton}(a) and displays exactly the conventional solitonic profile along the $y$-direction, see the inset of Fig. 4(b). The dependence of the soliton width with the peak {\it E}-field amplitude rendered in Fig. 4(b) is very similar to that found for TE optical solitons, predicting the existence of subwavelength optical solitons also for this polarization.

In conclusion, we have demonstrated that graphene monolayers can support both TE and TM spatial optical solitons due to the extremely large magnitude of its nonlinear third-order susceptibility. Moreover, we have shown that for feasible values of the input power these quasi-one dimensional optical solitons can have a subwavelength lateral width. We have also developed a quasi-analytical model that has a semi-quantitative value and that is able to predict the field intensities needed for soliton formation.  The existence of subwavelength optical solitons adds a new capability to the already broad range of optical phenomena associated with graphene structures.

\begin{acknowledgments}
This work has been funded by the Spanish Ministry of Science and Innovation under contracts MAT2008-06609-C02 and CSD2007-046-NanoLight.es and grants RyC-2009-05489 and JCI-2008-3123.
\end{acknowledgments}



\begin{thebibliography}{99}
\bibitem{comsol} We have used the implementation of the finite element method provided by the commercial software COMSOL Multiphysics.
\end{thebibliography}
\end{document}